\documentclass[aps,pra,twocolumn,floatfix,groupedaddress]{revtex4}
\usepackage{epsfig}
\usepackage{graphicx}
\usepackage{dcolumn}
\usepackage{longtable}
\usepackage{rotating}
\usepackage{pdfpages}
\usepackage{amsthm,amsmath}

\usepackage[colorlinks=true,linkcolor=blue,citecolor=red,urlcolor=magenta]{hyperref}

\usepackage[utf8]{inputenc}

\begin{document}
%\preprint{\today}

\title{Relativistic Coupled-cluster Theory Analysis of Properties of Co-like Ions}
%\vspace{0.5cm}

\author{Dillip K. Nandy$^1$\footnote{Email: nandy@ibs.re.kr}}
\affiliation{$^1$Center for Theoretical Physics of Complex Systems, Institute for Basic Science (IBS), Daejeon 34126, Korea}

\author{B. K. Sahoo$^2$\footnote{Email: bijaya@prl.res.in}}
\affiliation{$^2$Atomic, Molecular and Optical Physics Division, Physical Research Laboratory, Ahmedabad-380009, India}
\date{Received date; Accepted date}

%\vskip1.0cm

\begin{abstract}
Ionization potentials, excitation energies, transition properties, and hyperfine structure constants of the low-lying $3p^6 3d^{9}  \ ^2D_{5/2}$, 
$3p^6 3d^{9}  \ ^2D_{3/2}$, $3p^5 3d^{10} \ ^2P_{3/2}$ and $3p^5 3d^{10}  \ ^2P_{1/2}$ atomic states of the Co-like highly-charged ions such as Y$^{12+}$, Zr$^{13+}$, Nb$^{14+}$, Mo$^{15+}$, Tc$^{16+}$, Ru$^{17+}$, Rh$^{18+}$, Pd$^{19+}$, Ag$^{20+}$ and Cd$^{21+}$ are investigated. The singles and doubles approximated relativistic coupled-cluster theory in the framework of one electron removal Fock-space formalism is employed over the Dirac-Hartree-Fock calculations to account for the electron correlation effects for determining the aforementioned properties. Higher-order relativistic corrections due to the Breit interaction and quantum electrodynamics effects in the evaluation of energies are also quantified explicitly. Our estimated values are compared with the other available theoretical calculations and experimental results, which are found to be in good agreement with each other.     
\end{abstract} 

\maketitle

\section{Introduction}

The spectroscopic study of highly charged ions (HCIs) of heavy and moderately heavy elements have been the subject of primary interest in many 
contemporary areas of theoretical and experimental research fields. This includes tokamak plasmas and other high-temperature-plasma devices 
\cite{Putterich, Yanagibayashi}, electron beam ion trap (EBIT) \cite{Sudkewer, Suckewer1, Biemont, Utter, Porto, Yu, Gillaspy}, stellarators 
\cite{Harte}, atomic clocks \cite{Nandy0, Yu1, Yu2, Yu3, Safronova1} and probing fundamental physics \cite{Safronova1, Berengut, Dzuba, Berengut2}. 
One of the important implications of these HCIs is the use of their forbidden transition lines in plasma diagnostics. For example, various visible 
or ultraviolet magnetic-dipole (M1) transition lines of Ti-like ions were analyzed for density diagnostics in hot plasmas since the pioneering work 
of Feldman et al. \cite{Feldman}. Furthermore, accurate measurements of wavelengths, excitation energies and other spectroscopic properties of these 
ions also drive various theoretical research areas of the HCIs; especially in analyzing the astrophysical and laboratory plasma. Besides the plasma 
diagnostics, high-precision calculations of different radiative properties of the HCIs play an important role in testing several {\it ab initio} 
theories of quantum many-body systems where the relativistic and bound quantum electrodynamic (QED) effects play crucial roles in explaining the 
experimental predictions. This is why both the forbidden and allowed transition properties of various HCIs have been investigated in many earlier studies by employing various relativistic methods (e.g. see \cite{Nandy1, Nandy2, Nandy3, Nandy4, Safronova1, Cheung, Berengut, Ralchenko, Ding}). 

In the present study, we have investigated various transition properties of the highly charged Co-like transition metal ions such as Y$^{12+}$, 
Zr$^{13+}$, Nb$^{14+}$, Mo$^{15+}$, Tc$^{16+}$, Ru$^{17+}$, Rh$^{18+}$, Pd$^{19+}$, Ag$^{20+}$ and Cd$^{21+}$. In particular, we have calculated the 
first four low-lying atomic states of these ions in the framework of four-component relativistic coupled-cluster (RCC) theory. The four low-lying 
states include the $3p^6 3d^{9}  \ ^2D_{5/2}$, $3p^6 3d^{9}  \ ^2D_{3/2}$, $3p^5 3d^{10}  \ ^2P_{3/2}$ and $3p^5 3d^{10}  \ ^2P_{1/2}$ states,
which are in fact, one electron less than the $[3p^6 3d^{10}]$ closed-shell configuration; i.e. from the ground state configuration of the 
Ni isoelectronic sequence ions. Thus, it is convenient to adopt a Fock-space approach to determine the wave functions of the above states 
by starting calculations for the $[3p^6 3d^{10}]$ configuration. 

On the experimental interest of the Co-like ions, there are already a few observations available for several Co-like ions. For instance, Suckewer 
et al. identified the M1 transition lines between the fine-structure splitting of the ground state configuration of the Co-like Mo and Zr ions in 
the Princeton Large Torus tokamak plasma \cite{Suckewer3}. Similarly, Prior identified forbidden transitions of Nb$^{14+}$ in the emission lines from the intense, continuous beams of metastable HCIs produced by an electron cyclotron resonance ion source \cite{Prior}. There are also a few 
experimental identifications of lines available for the allowed $3p^{6}3d^9 \ ^2D_{5/2, 3/2} \rightarrow 3p^{5}3d^{10} \ ^2D_{1/2, 3/2}$ 
transitions. Edl\'{e}n first observed the allowed $3p^{6}3d^9 \rightarrow 3p^{5}3d^{10}$ transitions in the Sr$^{11+}$, Y$^{12+}$, Zr$^{13+}$, and
Mo$^{15+}$ HCIs in the spectra of hot tokamak plasmas along with other isoelectronic series of ions. Although, his observation did not yield any 
direct measurements of wavelengths for the Co-like ions as clearly made for the other isoelectronic series, however, it provided significant useful
information in identifying the allowed transition lines \cite{Edlen}. Ekberg et al. observed various electric-dipole (E1) transitions such as 
the $3p^{5}3d^{10} \ ^2P_{3/2} \rightarrow 3p^{6}3d^9 \ ^2D_{3/2}$, $3p^{5}3d^{10} \ ^2P_{5/2} \rightarrow 3p^{6}3d^9 \ ^2D_{3/2}$ and 
$3p^{5}3d^{10} \ ^2P_{1/2} \rightarrow 3p^{6}3d^9 \ ^2D_{3/2}$ transitions in Ru$^{17+}$, Rh$^{18+}$, Pd$^{19+}$, Ag$^{20+}$ and Cd$^{21+}$ along
with several other Co-like ions \cite{Ekberg}. Alexander et al. also reported measurements of these allowed transitions among the ground and first
excited states doublets of the Y$^{12+}$-Mo$^{15+}$ ions \cite{Alexander}. In another experiment, Burkhalter et al. observed the spectra of the 
Co-like Sr$^{11+}$, Y$^{12+}$, Zr$^{13+}$, Nb$^{14+}$, and Mo$^{15+}$ ions by employing a low inductance vacuum spark and a 10.7-m 
grazing-incidence spectrograph in the region $40-95$ {\AA} \cite{Burkhalter}. 

There are also a few theoretical calculations available on a number of Co-like ions but focusing mainly on the ground state fine structure 
splitting. For example, Guo et al. calculated the $3p^{6}3d^{9} \ ^2D_{5/2}$ and $3p^{6}3d^{9} \ ^2D_{3/2}$ states using the 
multi-configuration-Dirac-Hartree-Fock (MCDHF) and relativistic many-body perturbation theory (RMBPT) \cite{Guo}. They also estimated other 
transition properties involving these two states from their calculations. Their results show that values from the the MCDHF method provides 
relatively more accurate calculations than those are obtained using the RMBPT method. In another study, Chen et al. used an older version of 
the MCDHF code by Grant et al. \cite{Grant1} for determining the wavelengths of the fine structure splitting of the ground state configuration in
Zr$^{13+}$, Nb$^{14+}$ and Mo$^{15+}$, which predicted larger values for the wavelengths than that were obtained using the MCDHF and RMBPT methods
\cite{Guo}. Since the truncated RCC theory includes electron correlation effects to all-orders over the finite-order RMBPT method and take care of 
the size-inconsistency issue over the approximated MCDHF method, the calculations employing the RCC methods are believed to offer more reliable 
results for the transition properties of the investigated Co-like ions. Moreover, we have accounted for contributions from the leading order QED 
corrections and the Breit interaction effects mediated by the transverse component of the virtual photon between the electrons that are typically 
significant in the HCIs. 

The present paper is organized as follows. In Sec. \ref{AtomHam}, we briefly describe the approximations made in the Hamiltonian to include various 
physical effects within the atomic systems and the mean-field method considered as the initial approximation to generate the single particle atomic 
orbitals. In Sec. \ref{RCC}, we discuss about the Fock-space based RCC theory that is employed to determine the energies and transition matrix 
elements of the aforementioned states of the Co-like HCIs. Then, we present the formulas used to estimate the transition probabilities, lifetimes and hyperfine structure constants of the atomic states in Sec. \ref{prop}. In Sec. \ref{Results}, we present the results and discuss them in comparison with the previously reported values before concluding the work. Unless stated otherwise, all the quantities are given in atomic units (a.u.).

\section{\label{AtomHam} Approximations in Atomic Hamiltonian}

The general relativistic many-body Hamiltonian that incorporates the usual longitudinal component of the Coulomb interactions between the electrons
in an atomic system is given by
\begin{eqnarray}
H_{DC} =\sum^N_i \left [ c\mbox{\boldmath$\alpha$}_i\cdot \textbf{p}_i+(\beta_i -1)c^2+
V_{nuc}(r_i) + \sum_{j>i} V_{C}(r_{ij}) \right ]. \nonumber \\
\end{eqnarray}
Here, the subscript `DC' refers to the short-hand notation for the Dirac-Coulomb Hamiltonian, the first term describes the kinetic energy part of the electrons, the second term denotes the rescaling of atomic Hamiltonian by subtracting 
the rest mass energy of the electron, third term $V_{nuc}(r_i)$ is the nuclear potential with Fermi type charge distribution and the last term is the
two-body Coulomb repulsion term between the electrons. $N$ is the total number of the electron in the system and $\mbox{\boldmath$\alpha$}_i$ and $\beta_i$ 
are the usual $4 \times 4$ Dirac matrices. 

\begin{table*}[t]
\caption{\label{IPs}The calculated IPs (in cm$^{-1}$) of the orbitals $3d_{5/2}, 3d_{3/2}, 3p_{3/2}$ and $3p_{1/2}$ from the Ni-like closed-shell 
configuration $3p^6 3d^{10}$ for obtaining the atomic states $3p^6 3d^{9} \ ^2D_{5/2, 3/2}$ and $3p^5 3d^{10} \ ^2P_{1/2, 3/2}$ of the Y$^{12+}$, 
Zr$^{13+}$, Nb$^{14+}$, Mo$^{15+}$, Tc$^{16+}$, Ru$^{17+}$, Rh$^{18+}$, Pd$^{19+}$, Ag$^{20+}$ and Cd$^{21+}$ ions. Contributions from the Breit, 
VP (Uehling+Wichmann Kroll) and SE (both from the electric and magnetic form factors) effects are given as $\Delta E_B$, $\Delta E_{VP}$ and 
$\Delta E_{SE}$ respectively. The results obtained using the DC Hamiltonian are also given at different level of approximations such as 
DHF, RMBPT(2) and RCCSD methods. The final values are obtained by adding the RCCSD values from the DC Hamiltonian and other relativistic 
corrections. Our final RCCSD values for the $3p^6 3d^{9}  \ ^2D_{5/2}$ state is compared with the NIST data \cite{NIST}.} 
\begin{ruledtabular}
\begin{center}
%\scalebox{1.1}{
\begin{tabular}{lccccccccc}
%\hline \\
\vspace{1.3mm} \\
  State&\multicolumn{3}{c}{DC} & $\Delta E_B$ &$\Delta E_{VP}$ & $\Delta E_{SE}$  &\multicolumn{1}{c}{Final} & NIST   \\
\cline{2-4}  \\
       &       DHF & RMBPT(2) & RCCSD  &  &  &   &   RCCSD &   \\ 
%\hline \\
\vspace{1.3mm}
\textbf{Y$^{12+}$} \\
\vspace{1.3mm}
                     $3p^6 3d^{9}  \ ^2D_{5/2}$    & 3035558 & 3006606 & 3015229 &$-323    $ &$-13 $ &$131  $ & 3015024(2000) & 3016800(2000) \\
\vspace{1.3mm}                                                                                                                
                     $3p^6 3d^{9}  \ ^2D_{3/2}$    & 3054084 & 3024292 & 3033125 &$-1041   $ &$-13 $ &$221  $ & 3032291(1940) &  \\
\vspace{1.3mm}
                     $3p^5 3d^{10} \ ^2P_{3/2}$    & 4207130 & 4150875 & 4162105 &$-2664   $ &$-17 $ &$-23  $ & 4159401(1607) & \\
\vspace{1.3mm}
                     $3p^5 3d^{10} \ ^2P_{1/2}$    & 4310330 & 4249939 & 4262156 &$-4372   $ &$-13 $ &$319  $ & 4258090(1730) & \\
        
\vspace{1.3mm}
\textbf{Zr$^{13+}$} \\
\vspace{1.3mm}
                     $3p^6 3d^{9}  \ ^2D_{5/2}$    & 3465335 & 3436867 & 3444895 &$-459    $ &$-15 $ &$104  $ &3444525(1800)& 3436000(21000)  \\
\vspace{1.3mm}
                     $3p^6 3d^{9}  \ ^2D_{3/2}$    & 3486867 & 3457540 & 3465772 &$-1267   $ &$-15 $ &$260  $ &3464749(1740)&   \\
\vspace{1.3mm}
                     $3p^5 3d^{10} \ ^2P_{3/2}$    & 4693932 & 4639466 & 4649668 &$-3036   $ &$-18 $ &$-131 $ &4646481(1540)&   \\
\vspace{1.3mm}
                     $3p^5 3d^{10} \ ^2P_{1/2}$    & 4811553 & 4753037 & 4764098 &$-4944   $ &$-14 $ &$501  $ &4759641(1580)&   \\                     
\vspace{1.3mm}

\textbf{Nb$^{14+}$} \\
\vspace{1.3mm}
                     $3p^6 3d^{9}  \ ^2D_{5/2}$    & 3920121 & 3892095 & 3899604 &$-615    $ &$-16 $  &$162 $ &3899135(1780)& 3892000(12000) \\
\vspace{1.3mm}
                     $3p^6 3d^{9}  \ ^2D_{3/2}$    & 3945009 & 3916101 & 3923810 &$-1521   $ &$-17 $  &$201 $ &3922473(1700)&  \\
\vspace{1.3mm}
                     $3p^5 3d^{10} \ ^2P_{3/2}$    & 5206119 & 5153167 & 5162532 &$-3444   $ &$-20 $  &$20  $ &5159088(1520)&   \\
\vspace{1.3mm}
                     $3p^5 3d^{10} \ ^2P_{1/2}$    & 5339672 & 5282675 & 5292826 &$-5565   $ &$-15 $  &$120 $ &5287367(1640)&   \\
\vspace{1.3mm}

\textbf{Mo$^{15+}$} \\
\vspace{1.3mm}
                     $3p^6 3d^{9}  \ ^2D_{5/2}$    & 4399869 & 4372259 & 4379311 &$-792    $ &$-19 $  &$187 $ &4378686(1660)& 4388000 (4000) \\
\vspace{1.3mm}
                     $3p^6 3d^{9}  \ ^2D_{3/2}$    & 4428487 & 4399970 & 4407218 &$-1803   $ &$-19 $  &$224 $ &4405619(1680)&    \\
\vspace{1.3mm}
                     $3p^5 3d^{10} \ ^2P_{3/2}$    & 5743654 & 5692024 & 5700689 &$-3886   $ &$-23 $  &$27  $ &5696807(1520)&    \\
\vspace{1.3mm}
                     $3p^5 3d^{10} \ ^2P_{1/2}$    & 5894759 & 5839047 & 5848448 &$-6237   $ &$-16 $  &$104 $ &5842298(1620)&    \\                     
\vspace{1.3mm}

\textbf{Tc$^{16+}$} \\
\vspace{1.3mm}
                     $3p^6 3d^{9}  \ ^2D_{5/2}$    & 4904541 & 4877323 & 4883968 &$-992    $ &$-21 $  &$237 $ &4883191(1600)& 4872000(21000)  \\
\vspace{1.3mm}
                     $3p^6 3d^{9}  \ ^2D_{3/2}$    & 4937289 & 4909137 & 4915976 &$-2115   $ &$-22 $  &$248 $ &4914087(1610)&   \\
\vspace{1.3mm}
                     $3p^5 3d^{10} \ ^2P_{3/2}$    & 6306523 & 6256065 & 6264131 &$-4365   $ &$-25 $  &$79  $ &6259819(1550)&   \\
\vspace{1.3mm}
                     $3p^5 3d^{10} \ ^2P_{1/2}$    & 6476911 & 6422307 & 6431074 &$-6964   $ &$-17 $  &$19  $ &6424111(1530)&   \\                     
\vspace{1.3mm}

\textbf{Ru$^{17+}$} \\
\vspace{1.3mm}
                     $3p^6 3d^{9}  \ ^2D_{5/2}$    & 5434085 & 5407228 & 5413510 &$-1214   $ &$-24 $  &$221 $ &5412492(1570)& 5404000(23000)  \\
\vspace{1.3mm}
                     $3p^6 3d^{9}  \ ^2D_{3/2}$    & 5471390 & 5443569 & 5450045 &$-2458   $ &$-25 $  &$352 $ &5447913(1560)&   \\
\vspace{1.3mm}
                     $3p^5 3d^{10} \ ^2P_{3/2}$    & 6894708 & 6845288 & 6852837 &$-4882   $ &$-29 $  &$-65 $ &6847859(1400)&   \\
\vspace{1.3mm}
                     $3p^5 3d^{10} \ ^2P_{1/2}$    & 7086232 & 7032581 & 7040804 &$ -7747  $ &$-19 $  &$365 $ &7033403(1560)&   \\                     
\vspace{1.3mm}

\textbf{Rh$^{18+}$} \\
\vspace{1.3mm}
                     $3p^6 3d^{9}  \ ^2D_{5/2}$    & 5988428 & 5961888 & 5967851 &$-1460   $ &$-27 $  &$232 $ &5966596(1500)& 5960000(24000)  \\
\vspace{1.3mm}
                     $3p^6 3d^{9}  \ ^2D_{3/2}$    & 6030747 & 6003212 & 6009366 &$-2833   $ &$-28 $  &$395 $ &6006900(1480)&   \\
\vspace{1.3mm}
                     $3p^5 3d^{10} \ ^2P_{3/2}$    & 7508181 & 7459673 & 7466773 &$-5439   $ &$-32 $  &$-108$ &7461193(1360)&   \\
\vspace{1.3mm}
                     $3p^5 3d^{10} \ ^2P_{1/2}$    & 7722818 & 7669976 & 7677730 &$-8589   $ &$-19 $  &$435 $ &7669557(1580)&   \\
\end{tabular}
%}
\end{center}
\end{ruledtabular}
\end{table*}

\begin{table*}[htbp!]
\caption{Contd...} 
\begin{ruledtabular}
\begin{center}
%\scalebox{1.1}{
\begin{tabular}{lcccccccc}
%\hline \\
\vspace{1.3mm} \\
  State&\multicolumn{3}{c}{DC} & $\Delta E_B$ &$\Delta E_{VP}$ & $\Delta E_{SE}$  &\multicolumn{1}{c}{Final} & NIST    \\
\cline{2-4} \\
       &       DHF& RMBPT(2)& RCCSD  &  &  &   &   RCCSD &     \\ 
%\hline \\
\vspace{1.3mm}
\textbf{Pd$^{19+}$} \\
\vspace{1.3mm}
                     $3p^6 3d^{9}  \ ^2D_{5/2}$    & 6567479 & 6541206 & 6546886 &$-1730  $ &$-30  $  &$226  $ &6545351(1580)& 6533000(25000)   \\
\vspace{1.3mm}
                     $3p^6 3d^{9}  \ ^2D_{3/2}$    & 6615298 & 6587997 & 6593866 &$-3241  $ &$-31  $  &$444  $ &6591037(1520)&    \\
\vspace{1.3mm}
                     $3p^5 3d^{10} \ ^2P_{3/2}$    & 8146906 & 8099189 & 8105897 &$-6036  $ &$-34  $  &$-191 $ &8099634(1430)&    \\
\vspace{1.3mm}
                     $3p^5 3d^{10} \ ^2P_{1/2}$    & 8386760 & 8334593 & 8341939 &$-9491  $ &$-20  $  &$565  $ &8332992(1540)&    \\
\vspace{1.3mm}

\textbf{Ag$^{20+}$} \\
\vspace{1.3mm}
                     $3p^6 3d^{9}  \ ^2D_{5/2}$    & 7171144 & 7145090 & 7150516 &$-2026  $ &$-33  $  &$-148  $ &7148307(1500)& 7138000(30000)  \\
\vspace{1.3mm}
                     $3p^6 3d^{9}  \ ^2D_{3/2}$    & 7224981 & 7197861 & 7203477 &$-3684  $ &$-33  $  &$391   $ &7200149(1460)&   \\
\vspace{1.3mm}
                     $3p^5 3d^{10} \ ^2P_{3/2}$    & 8810843 & 8763808 & 8770169 &$-6676  $ &$-37  $  &$-871  $ &8762585(1440)&   \\
\vspace{1.3mm}
                     $3p^5 3d^{10} \ ^2P_{1/2}$    & 9078153 & 9026539 & 9033526 &$-10457 $ &$-21  $  &$1413  $ &9024461(1560)&   \\                     
\vspace{1.3mm}

\textbf{Cd$^{21+}$} \\
\vspace{1.3mm}
                     $3p^6 3d^{9}  \ ^2D_{5/2}$    & 7799345 & 7773465 & 7778662 &$-2350  $ &$-36  $  &$304   $ &7776580(1580)& 7767000(30000)   \\
\vspace{1.3mm}
                     $3p^6 3d^{9}  \ ^2D_{3/2}$    & 7859747 & 7832764 & 7838151 &$-4164  $ &$-37  $  &$437   $ &7834387(1540)&    \\
\vspace{1.3mm}
                     $3p^5 3d^{10} \ ^2P_{3/2}$    & 9499967 & 9453517 & 9459572 &$-7359  $ &$-40  $  &$-109  $ &9452061(1500)&    \\
\vspace{1.3mm}
                     $3p^5 3d^{10} \ ^2P_{1/2}$    & 9797108 & 9745940 & 9752610 &$-11489 $ &$-21  $  &$231   $ &9741331(1750)&    \\
\vspace{1.3mm}

\end{tabular}
%}
\end{center}
\end{ruledtabular}
\end{table*}

Since the considered systems are highly charged, so the relativistic effects in these ions are anticipated to be quite large. Therefore, for the 
accurate calculations of excitation spectra and transitions properties, it is necessary to incorporate higher-order relativistic effects at the 
one-body and two-body levels. At the two-body level, higher-order relativistic effects are accounted through the Breit-interactions mediated 
by the exchange of the transverse component of the virtual photon between the electrons and have the form \cite{Grant}
\begin{eqnarray}
V_{Brt}(r_{ij})=- \frac{1}{2r_{ij}}\{\mbox{\boldmath$\alpha$}_i\cdot \mbox{\boldmath$\alpha$}_j+
(\mbox{\boldmath$\alpha$}_i\cdot\bf{\hat{r}_{ij}})(\mbox{\boldmath$\alpha$}_j\cdot\bf{\hat{r}_{ij}}) \}, 
\end{eqnarray}
where $r_{ij} = |\vec{r_i} - \vec{r_j}|$ denotes the absolute magnitude of the difference between radial vectors of any two electrons at positions 
$\vec r_i$ and $\vec{r_j}$. Similarly, the higher-order relativistic effects that occur between the electrons and the nucleus is taken into the 
nuclear potential energy by defining effective model potentials. This includes leading order vacuum-polarization (VP) and self-energy (SE) effects. 
In our calculation, the net effective QED potential of an electron at the position $r_i$ is expressed as
\begin{eqnarray}
V^{QED}_{nuc}(r_i) = V_{Uhl}(r_i)+\frac{2}{3}V^{simple}_{WK}(r_i) + V_{mf}(r_i) + V_{ef}(r_i). \nonumber \\
\end{eqnarray}
The first two terms $V_{Uhl}(r_i)$ and $V^{simple}_{WK}(r_i)$ are known as the Uehling and Wichmann-Kroll model potentials arising due to the VP effects on the bound electrons. Similarly, the last two terms $V_{mf}(r_i)$ and $V_{ef}(r_i)$ represent the magnetic and electric form factors arising due to the SE corrections to the bound electrons. Analytical expressions for these $V_{Uhl}(r_i)$, $V^{simple}_{WK}(r_i)$, $V_{mf}(r_i)$ and $V_{ef}(r_i)$ terms are given by \cite{Ginges, Sahoo2}
\begin{eqnarray}
V_{Uhl}(r)&=&  - \frac{4 \alpha^2 }{9r} V_{fermi}(r) \nonumber \\
&&\times \int_1^{\infty}dt\sqrt{t^2-1}\left(\frac{1}{t^2}+\frac{1}{2t^4}\right)e^{-2r t/ \alpha } ,
\end{eqnarray}
\begin{eqnarray}
V^{simple}_{WK}(r)=-\frac{2}{3}\frac{\alpha}{\pi}V_{fermi}(r)\frac{0.092Z^2/\alpha^2}{1+(1.62r/\alpha)^4} ,
\end{eqnarray}
\begin{eqnarray}
V_{mf}(r)=\frac{\alpha^2}{4\pi}i \mbox{\boldmath$\vec{\gamma}$}.\mbox{\boldmath$\vec{\bigtriangledown}$} 
\left [V_{fermi}(r)\left (\int^{\infty}_{1}dt \frac{1}{\sqrt{t^2-1}}e^{-2tr/\alpha}\right)\right]\nonumber \\ 
\end{eqnarray}
and 
\begin{eqnarray}
V_{ef}(r)&=&-A(Z,r)\frac{\alpha}{\pi}V_{fermi}(r)\int^{\infty}_1 dt \frac{e^{-2tr/\alpha}}{\sqrt{t^2-1}}
[\left( 1-\frac{1}{2t^2}\right)\nonumber \\
&&\times \{ln(t^2-1)+4 ln(1/Z\alpha+0.5)\}-\frac{3}{2}+\frac{1}{t^2}] \nonumber \\
&&\times B(Z)Z^4\alpha^3 e^{-Zr}, 
\end{eqnarray}
where the factors $A(Z,r)=[1.071-1.97((Z-80)\alpha)^2-2.128((Z-80)\alpha)^3+0.169((Z-80)\alpha)^4](r/\alpha)(r/\alpha+0.07Z^2\alpha^2)$ and 
$B(Z)=0.074+0.35Z\alpha$.

Thus, the final Hamiltonian that has been used in the present calculation has the following form
\begin{eqnarray}
H_{DCBVS} = H_{DC} + \sum^N_i \left [V^{QED}_{nuc}(r_i) + \sum_{j>i} V_{Brt}(r_{ij}) \right ]. \label{H_DCBVS} 
\end{eqnarray}
The exact solution of the above Hamiltonian is not possible due to the two-body interaction terms (Coulomb and Breit), so one of the practical 
approaches to tackle the many-body problem is to start with a mean-field approximation. In the present work, we use the relativistic Hartree-Fock (HF) or Dirac-Hartree-Fock (DHF) method to obtain the mean-field wave function $|\Phi_0 \rangle$ of the $[3p^6 3d^{10}]$ closed-shell configuration, 
its detail underlying theory can be found elsewhere \cite{Lindgren, Reiher, Johnson}, to obtain the single-particle orbitals of the considered atomic systems. 
    
To carry out the calculations conveniently, we define the normal order form of the atomic Hamiltonian defined with respect to the (D)HF wave function $|\Phi_0 \rangle$ (reference state) of the $[3p^6 3d^{10}]$ closed-shell configuration in this case by defining
\begin{eqnarray}
H_N &=& H_{DCBVS}-\langle\Phi_0|H_{DCBVS}|\Phi_0\rangle \nonumber \\
    &=& H_{DCBVS} - E_{SCF}, \label{Hnorm}
\end{eqnarray}
with the self-consistent-field (SCF) energy $E_{SCF}$. Then, we employ the Fock-space approach to obtain the atomic wave functions of the 
$3p^6 3d^{9}  \ ^2D_{5/2}$, $3p^6 3d^{9}  \ ^2D_{3/2}$, $3p^5 3d^{10}  \ ^2P_{3/2}$ and $3p^5 3d^{10}  \ ^2P_{1/2}$ states of the Co-like ions.

\begin{table*}[t]
\caption{\label{EEs}Comparison of our calculated EEs (in cm$^{-1}$) of the low-lying excited $3p^6 3d^{9} \ ^2D_{3/2}$, $3p^5 3d^{10} \ 
^2P_{3/2}$ and $3p^5 3d^{10} \ ^2P_{1/2}$ states of the considered Co-like Y$^{12+}$ - Cd$^{21+}$ ions. The values indicated under `Present' 
are deducted from the differences of our calculated IPs given in the previous table while direct measured values \cite{Edlen,Prior} are quoted as 
`Experiment'. The values given under `Fitted' are the extrapolated values reported in the literature by combining calculations using the 
MCDHF method and the observed wavelength values \cite{Ekberg}.}
\begin{ruledtabular}
\begin{center}
\scalebox{1.0}{
\begin{tabular}{lcccccccccccc}
%\hline \\
\vspace{1.3mm} \\
  Ion & \multicolumn{1}{c}{$3p^6 3d^{9}  \ ^2D_{5/2}$}  &  \multicolumn{3}{c}{$3p^6 3d^{9}  \ ^2D_{3/2}$}&  &\multicolumn{2}{c}{$3p^5 3d^{10}  \ ^2P_{3/2}$}& &\multicolumn{2}{c}{$3p^5 3d^{10} \ ^2P_{1/2}$}  \\
\cline{3-5} \cline{7-8} \cline{10-11}  \\
                   &     & Present  & Experiment   & Fitted$^c$ &   &Present & Fitted$^c$ &  & Present  & Fitted$^c$                      \\   \\
%\hline \\
\vspace{1.3mm}
\text{Y$^{12+}$}      &0.0  &17267  &              &$17240(10) $ &   &1144377  & $1144220(70)$ &  &1243066 &$1242580(80) $     \\
\vspace{1.3mm}                  
\text{Zr$^{13+}$}     &0.0  &20224  &$20131(1.0)^a$&$20125(1.2)$ &   &1201956  & $1201940(70)$ &  &1315116 &$1314590(80) $    \\
\vspace{1.3mm}                  
\text{Nb$^{14+}$}     &0.0  &23338  &$23369(5.0)^b$&$23363(5)$   &   &1259953  & $1259890(80)$ &  &1388232 &$1388250(90) $    \\
\vspace{1.3mm}                  
\text{Mo$^{15+}$}     &0.0  &26933  &$26967(2.0)^a$&$26960(1.5)$ &   &1318121  & $1318110(90)$ &  &1463612 &$1463760(100)$    \\
\vspace{1.3mm}                                        
\text{Tc$^{16+}$}     &0.0  &30896  &              &$30950(30) $ &   &1376628  & $1376670(90)$ &  &1540920 &$1541270(120)$    \\
\vspace{1.3mm}                                             
\text{Ru$^{17+}$}     &0.0  &35421  &              &$35360(40) $ &   &1435367  & $1435610(100)$&  &1620911 &$1621000(130)$    \\
\vspace{1.3mm}                                             
\text{Rh$^{18+}$}     &0.0  &40304  &              &$40230(40) $ &   &1494597  & $1494970(110)$&  &1702961 &$1703130(140)$    \\
\vspace{1.3mm}                                             
\text{Pd$^{19+}$}     &0.0  &45686  &              &$45580(50) $ &   &1554283  & $1554850(120)$&  &1787641 &$1787800(160)$    \\
\vspace{1.3mm}                  
\text{Ag$^{20+}$}     &0.0  &51842  &              &$51430(50) $ &   &1614278  & $1615220(130)$&  &1876154 &$1875220(170)$    \\
\vspace{1.3mm}                  
\text{Cd$^{21+}$}     &0.0  &57807  &              &$57810(60) $ &   &1675481  & $1676160(140)$&  &1964751 &$1965520(190)$    \\
\end{tabular}
}
\end{center}
\end{ruledtabular}
\begin{tabular}{lc}
References: 
$^a$\cite{Suckewer3}; 
$^b$\cite{Prior};
$^c$\cite{Ekberg}.  
\end{tabular}
\end{table*}

\section{\label{RCC}RCC method for one electron detachment}

As mentioned earlier, the atomic states that are being investigated in the reported HCIs are the four low-lying states of the Co isoelectronic 
series, which are the $3p^6 3d^{9} \ ^2D_{5/2}$, $3p^6 3d^{9} \ ^2D_{3/2}$, $3p^5 3d^{10} \ ^2P_{3/2}$, $3p^5 3d^{10} \ ^2P_{1/2}$ states, and 
their configurations are one electron short of the closed-shell configuration $[3p^6 3d^{10}]$. We consider here single-referee RCC theory in the 
similar philosophy of electron detachment approach as discussed in \cite{Bartlett, Lindgren} to obtain the wave functions of the above states. The 
basic strategy of this approach is described briefly as follows. After obtaining the DHF wave function $|\Phi_0 \rangle$ of the $[3p^6 3d^{10}]$ 
closed-shell configuration, we determine its exact wave function using the RCC theory ansatz \cite{Lindgren, Bartlett}
\begin{eqnarray}
 |\Psi_0 \rangle = e^T |\Phi_0 \rangle ,
\end{eqnarray}
where $T$ is defined as the linear combinations of all possible hole-particle excitation operators that are responsible for accounting the 
neglected residual interactions in the calculation of the DHF wave function. The amplitudes of these operators are obtained by solving the non-linear
equation \cite{Lindgren, Bartlett, Sahoo1}
\begin{eqnarray}
\langle \Phi^*_0| \widehat{H_N e^T}|\Phi_0 \rangle &=& 0, \label{T0eng}
\end{eqnarray}
where $|\Phi_0^* \rangle$ represents for the excited Slater determinants with respect to $|\Phi_0 \rangle$. After obtaining the RCC amplitudes, 
the exact energy of the $[3p^6 3d^{10}]$ configuration is obtained by 
\begin{eqnarray}
 E_0 = E_{SCF} + \langle \Phi_0 | H_N | \Phi_0 \rangle .
\end{eqnarray} 

\begin{table*}[t]
%\begin{sideways}
%\centering
\caption{\label{Prob} Transition properties such as line strengths $S^O_{if}$ (in a.u.), oscillator strengths $F_{if}^O$ and transition rates 
$A^O_{if}$ (in $s^{-1}$) due to different channels ($O$) for the five low-lying transitions among the atomic states calculated in this work. The 
values obtained using our RCCSD method are denoted as `This work' and they are compared with the previously reported values using the MCDHF method 
\cite{Guo}. The estimated lifetimes $\tau_i$ (in $s$) for the excited atomic states $3p^{6}3d^9  \ ^2D_{3/2}$, $3p^{5}3d^{10}  \ ^2P_{3/2}$ and 
$3p^{5}3d^{10} \ ^2P_{1/2}$ using the total transition probabilities are listed from both the works are listed in the last two columns.}
\begin{ruledtabular}
\begin{center}
\scalebox{0.95}{
\begin{tabular}{lcccccccccc}          
Transition ($O$) & \multicolumn{2}{c}{$S_{if}^O$}  & \multicolumn{2}{c}{$F_{if}^O$}& \multicolumn{2}{c}{$A_{if}^O$}  &  \multicolumn{2}{c}{$\tau_i$} \\ 
\cline{2-3} \cline{4-5} \cline{6-7} \cline{8-9} \\
            & This work & Ref. \cite{Guo}   & This work & Ref. \cite{Guo}   & This work  & Ref. \cite{Guo}   & This work  & Ref. \cite{Guo}   \\
\hline \\
\textbf{Y$^{12+}$} & & \\
\vspace{1.0mm}
$3p^{6}3d^9    \ ^2D_{3/2} \xrightarrow{M1} 3p^{6}3d^9    \ ^2D_{5/2}$    & 2.541  &2.395 &   2.952[-7]  &1.670[-6] &   87.80     &82.72    & 1.139[-2]  & 1.21[-2]   \\
\vspace{1.0mm}       
$~~~~~~~~~~~~~~~~~~        \xrightarrow{E2} 3p^{6}3d^9    \ ^2D_{5/2}$    & 0.0198 & &   2.842[-12] & &   8.452[-4]  &    &            &    \\

\vspace{1.0mm}       
$3p^{5}3d^{10} \ ^2P_{3/2} \xrightarrow{E1} 3p^{6}3d^9    \ ^2D_{3/2}$    & 0.0387 & &   0.039      & &   2.810[10]  &    & 3.380[-12] &     \\ 
\vspace{1.0mm}
$~~~~~~~~~~~~~~~~~~        \xrightarrow{E1} 3p^{6}3d^9    \ ^2D_{5/2}$    & 0.352  & &   0.203      & &   2.678[11]  &    &            &     \\

\vspace{1.0mm}       
$3p^{5}3d^{10} \ ^2P_{1/2} \xrightarrow{M1} 3p^{5}3d^{10} \ ^2P_{3/2}$    & 1.490  & &   1.482[-6]  & &   1.912[4]   &    & 2.800[-12] &     \\ 
\vspace{1.0mm}
$~~~~~~~~~~~~~~~~~~        \xrightarrow{E2} 3p^{5}3d^{10} \ ^2P_{3/2}$    & 0.0429 & &   2.014[-9]  & &   26.00      &    &            &     \\
\vspace{1.0mm}
$~~~~~~~~~~~~~~~~~~        \xrightarrow{E1} 3p^{6}3d^9    \ ^2D_{3/2}$    & 0.191  & &   0.177      & &   3.571[11]  &    &            &     \\

\textbf{Zr$^{13+}$} & & \\
\vspace{1.0mm}
$3p^{6}3d^9    \ ^2D_{3/2} \xrightarrow{M1} 3p^{6}3d^9    \ ^2D_{5/2}$    & 2.529  &2.395 &   3.431[-7]  &1.670[-6] &   139.12        &131.6    &  7.187[-3] &7.60[-3]     \\
\vspace{1.0mm}       
$~~~~~~~~~~~~~~~~~~        \xrightarrow{E2} 3p^{6}3d^9    \ ^2D_{5/2}$    & 0.0163 & &   3.733[-12] & &   0.0015     &    &            &     \\

\vspace{1.0mm}       
$3p^{5}3d^{10} \ ^2P_{3/2} \xrightarrow{E1} 3p^{6}3d^9    \ ^2D_{3/2}$    & 0.0357 & &   0.0317     & &   2.986[10]  &    & 3.164[-12] &     \\ 
\vspace{1.0mm}
$~~~~~~~~~~~~~~~~~~        \xrightarrow{E1} 3p^{6}3d^9    \ ^2D_{5/2}$    & 0.325  & &   0.198      & &   2.862[11] &    &            &     \\

\vspace{1.0mm}       
$3p^{5}3d^{10} \ ^2P_{1/2} \xrightarrow{M1} 3p^{5}3d^{10} \ ^2P_{3/2}$    & 1.466  & &   1.670[-6]  & &   2.828[4]   &    & 2.585[-12] &      \\ 
\vspace{1.0mm}
$~~~~~~~~~~~~~~~~~~        \xrightarrow{E2} 3p^{5}3d^{10} \ ^2P_{3/2}$    & 0.0429 & &   2.574[-9]  & &   43.569     &    &            &      \\
\vspace{1.0mm}
$~~~~~~~~~~~~~~~~~~        \xrightarrow{E1} 3p^{6}3d^9    \ ^2D_{3/2}$    & 0.176  & &   0.172      & &   3.868[11]  &    &            &      \\

\textbf{Nb$^{14+}$} & & \\
\vspace{1.0mm}
$3p^{6}3d^9    \ ^2D_{3/2} \xrightarrow{M1} 3p^{6}3d^9    \ ^2D_{5/2}$    & 2.518  &2.395 &   3.966[-7]  &2.263[-6] &   216.71     &205.7    & 4.614[-3]  &4.86[-3]      \\
\vspace{1.0mm}       
$~~~~~~~~~~~~~~~~~~        \xrightarrow{E2} 3p^{6}3d^9    \ ^2D_{5/2}$    & 0.0136 & &   4.867[-12] & &   0.0266     &    &            &      \\

\vspace{1.0mm}       
$3p^{5}3d^{10} \ ^2P_{3/2} \xrightarrow{E1} 3p^{6}3d^9    \ ^2D_{3/2}$    & 0.0331 & &   0.031      & &   3.168[10]  &    & 2.971[-12] &      \\ 
\vspace{1.0mm}
$~~~~~~~~~~~~~~~~~~        \xrightarrow{E1} 3p^{6}3d^9    \ ^2D_{5/2}$    & 0.301  & &   0.192      & &   3.052[11]  &    &            &      \\

\vspace{1.0mm}       
$3p^{5}3d^{10} \ ^2P_{1/2} \xrightarrow{M1} 3p^{5}3d^{10} \ ^2P_{3/2}$    & 1.449  & &   1.880[-6]  & &   4.133[4]   &    & 2.386[-12] &      \\ 
\vspace{1.0mm}
$~~~~~~~~~~~~~~~~~~        \xrightarrow{E2} 3p^{5}3d^{10} \ ^2P_{3/2}$    & 0.0368 & &   3.264[-9]  & &   71.745     &    &            &      \\
\vspace{1.0mm}
$~~~~~~~~~~~~~~~~~~        \xrightarrow{E1} 3p^{6}3d^9    \ ^2D_{3/2}$    & 0.162  & &   0.167      & &   4.190[11]  &    &            &      \\

\textbf{Mo$^{15+}$} & & \\
\vspace{1.0mm}
$3p^{6}3d^9    \ ^2D_{3/2} \xrightarrow{M1} 3p^{6}3d^9    \ ^2D_{5/2}$    & 2.508  &2.394 &   4.559[-7]  &2.610[-6] &   331.74    &316.3    & 3.014[-3]  &3.16[-3]      \\
\vspace{1.0mm}       
$~~~~~~~~~~~~~~~~~~        \xrightarrow{E2} 3p^{6}3d^9    \ ^2D_{5/2}$    & 0.0115 & &   6.290[-12] & &   0.0046     &    &            &      \\

\vspace{1.0mm}       
$3p^{5}3d^{10} \ ^2P_{3/2} \xrightarrow{E1} 3p^{6}3d^9    \ ^2D_{3/2}$    & 0.0307 & &   0.032      & &   3.349[10]  &    & 2.793[-12] &      \\ 
\vspace{1.0mm}
$~~~~~~~~~~~~~~~~~~        \xrightarrow{E1} 3p^{6}3d^9    \ ^2D_{5/2}$    & 0.279  & &   0.184      & &   3.247[11]  &    &            &      \\

\vspace{1.0mm}       
$3p^{5}3d^{10} \ ^2P_{1/2} \xrightarrow{M1} 3p^{5}3d^{10} \ ^2P_{3/2}$    & 1.435  & &   2.113[-6]  & &   5.980[4]   &    & 2.204[-12] &       \\ 
\vspace{1.0mm}
$~~~~~~~~~~~~~~~~~~        \xrightarrow{E2} 3p^{5}3d^{10} \ ^2P_{3/2}$    & 0.0317 & &   4.114[-9]  & &   116.431    &    &            &       \\
\vspace{1.0mm}
$~~~~~~~~~~~~~~~~~~        \xrightarrow{E1} 3p^{6}3d^9    \ ^2D_{3/2}$    & 0.150  & &   0.164      & &   4.536[11]  &    &            &       \\

\textbf{Tc$^{16+}$} & & \\
\vspace{1.0mm}
$3p^{6}3d^9    \ ^2D_{3/2} \xrightarrow{M1} 3p^{6}3d^9    \ ^2D_{5/2}$    & 2.500  &2.394 &   5.214[-7]  &2.996[-6] &   500.00    &478.5    & 2.001[-3]  &2.09[-3]       \\
\vspace{1.0mm}       
$~~~~~~~~~~~~~~~~~~        \xrightarrow{E2} 3p^{6}3d^9    \ ^2D_{5/2}$    & 0.010  & &   8.061[-12] & &   0.0077     &    &            &       \\

\vspace{1.0mm}       
$3p^{5}3d^{10} \ ^2P_{3/2} \xrightarrow{E1} 3p^{6}3d^9    \ ^2D_{3/2}$    & 0.0286 & &   0.030      & &   3.533[10]  &    & 2.631[-12] &       \\ 
\vspace{1.0mm}
$~~~~~~~~~~~~~~~~~~        \xrightarrow{E1} 3p^{6}3d^9    \ ^2D_{5/2}$    & 0.261  & &   0.181      & &   3.448[11]   &    &            &       \\

\vspace{1.0mm}       
$3p^{5}3d^{10} \ ^2P_{1/2} \xrightarrow{M1} 3p^{5}3d^{10} \ ^2P_{3/2}$    & 1.424  & &   2.369[-6]  & &   8.562[4]   &    & 2.038[-12] &       \\ 
\vspace{1.0mm}
$~~~~~~~~~~~~~~~~~~        \xrightarrow{E2} 3p^{5}3d^{10} \ ^2P_{3/2}$    & 0.0275 & &   5.122[-9]  & &   185.214    &    &            &       \\
\vspace{1.0mm}
$~~~~~~~~~~~~~~~~~~        \xrightarrow{E1} 3p^{6}3d^9    \ ^2D_{3/2}$    & 0.140  & &   0.160      & &   4.907[11]  &    &            &       \\

\end{tabular}
}
\end{center}
\end{ruledtabular}
\end{table*}

\begin{table*}[htbp!]
%\begin{sideways}
%\centering
\caption{\label{prob} Contd...}
\begin{ruledtabular}
\begin{center}
\scalebox{0.95}{
\begin{tabular}{lcccccccccc}          
Transition ($O$) & \multicolumn{2}{c}{$S_{if}^O$}  & \multicolumn{2}{c}{$F_{if}^O$}& \multicolumn{2}{c}{$A_{if}^O$}  &  \multicolumn{2}{c}{$\tau_i$} \\ 
\cline{2-3} \cline{4-5} \cline{6-7} \cline{8-9} \\
            & This work & Ref. \cite{Guo}  & This work & Ref. \cite{Guo}   & This work  & Ref. \cite{Guo}   & This work  & Ref. \cite{Guo} \\
\hline \\
\textbf{Ru$^{17+}$} & & \\
\vspace{1.0mm}
$3p^{6}3d^9    \ ^2D_{3/2} \xrightarrow{M1} 3p^{6}3d^9    \ ^2D_{5/2}$ & 2.492   &2.393   &  5.938[-7]   &3.422[-6]    & 742.92    &713.6   & 1.346[-3] &1.40[-3]   \\
\vspace{1.0mm}       
$~~~~~~~~~~~~~~~~~~        \xrightarrow{E2} 3p^{6}3d^9    \ ^2D_{5/2}$ & 0.0083  &   &  1.027[-11]  &    & 0.0130    &   &           &   \\

\vspace{1.0mm}       
$3p^{5}3d^{10} \ ^2P_{3/2} \xrightarrow{E1} 3p^{6}3d^9    \ ^2D_{3/2}$ & 0.0267  &   &  0.0281      &    & 3.717[10] &   & 2.484[-12]&   \\ 
\vspace{1.0mm}
$~~~~~~~~~~~~~~~~~~        \xrightarrow{E1} 3p^{6}3d^9    \ ^2D_{5/2}$ & 0.244   &   &  0.177       &    & 3.654[11] &   &           &   \\

\vspace{1.0mm}       
$3p^{5}3d^{10} \ ^2P_{1/2} \xrightarrow{M1} 3p^{5}3d^{10} \ ^2P_{3/2}$ & 1.414   &   &  2.650[-6]   &    & 1.215[5]  &   & 1.886[-12]&   \\ 
\vspace{1.0mm}
$~~~~~~~~~~~~~~~~~~        \xrightarrow{E2} 3p^{5}3d^{10} \ ^2P_{3/2}$ & 0.0240  &   &  6.423[-9]   &    & 294.50    &   &           &   \\
\vspace{1.0mm}
$~~~~~~~~~~~~~~~~~~        \xrightarrow{E1} 3p^{6}3d^9    \ ^2D_{3/2}$ & 0.131   &   &  0.157       &    & 5.303[11] &   &           &   \\

\textbf{Rh$^{18+}$} & & \\
\vspace{1.0mm}
$3p^{6}3d^9    \ ^2D_{3/2} \xrightarrow{M1} 3p^{6}3d^9    \ ^2D_{5/2}$ & 2.485   &2.393   &  6.737[-7]   &3.892[-6]    & 1090.92   &1050   & 9.166[-4] &9.53[-4]   \\
\vspace{1.0mm}       
$~~~~~~~~~~~~~~~~~~        \xrightarrow{E2} 3p^{6}3d^9    \ ^2D_{5/2}$ & 0.0071  &   &  1.300[-11]  &    & 0.0210    &   &           &   \\

\vspace{1.0mm}       
$3p^{5}3d^{10} \ ^2P_{3/2} \xrightarrow{E1} 3p^{6}3d^9    \ ^2D_{3/2}$ & 0.0250  &   &  0.0276      &    & 3.903[10] &   & 2.349[-12]&   \\ 
\vspace{1.0mm}
$~~~~~~~~~~~~~~~~~~        \xrightarrow{E1} 3p^{6}3d^9    \ ^2D_{5/2}$ & 0.228   &   &  0.173       &    & 3.866[11] &   &           &   \\

\vspace{1.0mm}       
$3p^{5}3d^{10} \ ^2P_{1/2} \xrightarrow{M1} 3p^{5}3d^{10} \ ^2P_{3/2}$ & 1.405   &   &  2.957[-6]   &    & 1.710[5]  &   & 1.746[-12]&   \\ 
\vspace{1.0mm}
$~~~~~~~~~~~~~~~~~~        \xrightarrow{E2} 3p^{5}3d^{10} \ ^2P_{3/2}$ & 0.0210  &   &  7.971[-9]   &    & 460.78    &   &           &   \\
\vspace{1.0mm}
$~~~~~~~~~~~~~~~~~~        \xrightarrow{E1} 3p^{6}3d^9    \ ^2D_{3/2}$ & 0.123   &   &  0.154       &    & 5.727[11] &   &           &   \\

\textbf{Pd$^{19+}$} & & \\
\vspace{1.0mm}
$3p^{6}3d^9    \ ^2D_{3/2} \xrightarrow{M1} 3p^{6}3d^9    \ ^2D_{5/2}$ & 2.478   &2.392   &  7.613[-7]   &4.408[-6]    & 1582.40   &1526   & 6.319[-4] &6.55[-4]   \\
\vspace{1.0mm}       
$~~~~~~~~~~~~~~~~~~        \xrightarrow{E2} 3p^{6}3d^9    \ ^2D_{5/2}$ & 0.0062  &   &  1.637[-11]  &    & 0.0340    &   &           &   \\

\vspace{1.0mm}       
$3p^{5}3d^{10} \ ^2P_{3/2} \xrightarrow{E1} 3p^{6}3d^9    \ ^2D_{3/2}$ & 0.0235  &   &  0.0267      &    & 4.088[10] &   & 2.224[-12]&   \\ 
\vspace{1.0mm}
$~~~~~~~~~~~~~~~~~~        \xrightarrow{E1} 3p^{6}3d^9    \ ^2D_{5/2}$ & 0.215   &   &  0.167       &    & 4.085[11] &   &           &   \\

\vspace{1.0mm}       
$3p^{5}3d^{10} \ ^2P_{1/2} \xrightarrow{M1} 3p^{5}3d^{10} \ ^2P_{3/2}$ & 1.398   &   &  1.578[-6]   &    & 2.624[5]  &   & 1.617[-12]&   \\ 
\vspace{1.0mm}
$~~~~~~~~~~~~~~~~~~        \xrightarrow{E2} 3p^{5}3d^{10} \ ^2P_{3/2}$ & 0.0185  &   &  9.889[-9]   &    & 718.45    &   &           &   \\
\vspace{1.0mm}
$~~~~~~~~~~~~~~~~~~        \xrightarrow{E1} 3p^{6}3d^9    \ ^2D_{3/2}$ & 0.1150  &   &  0.152       &    & 6.182[11] &   &           &   \\

\textbf{Ag$^{20+}$} & & \\
\vspace{1.0mm}
$3p^{6}3d^9    \ ^2D_{3/2} \xrightarrow{M1} 3p^{6}3d^9    \ ^2D_{5/2}$ & 2.472   &2.392   &  8.569[-7]   &4.972[-6]    & 2267.84   &2191   & 4.409[-4] &4.56[-4]   \\
\vspace{1.0mm}       
$~~~~~~~~~~~~~~~~~~        \xrightarrow{E2} 3p^{6}3d^9    \ ^2D_{5/2}$ & 0.0053  &   &  2.047[-11]  &    & 0.0542    &   &           &   \\

\vspace{1.0mm}       
$3p^{5}3d^{10} \ ^2P_{3/2} \xrightarrow{E1} 3p^{6}3d^9    \ ^2D_{3/2}$ & 0.0221  &   &  0.026       &    & 4.275[10] &   & 2.112[-12]&    \\ 
\vspace{1.0mm}
$~~~~~~~~~~~~~~~~~~        \xrightarrow{E1} 3p^{6}3d^9    \ ^2D_{5/2}$ & 0.202   &   &  0.165       &    & 4.309[11] &   &           &    \\

\vspace{1.0mm}       
$3p^{5}3d^{10} \ ^2P_{1/2} \xrightarrow{M1} 3p^{5}3d^{10} \ ^2P_{3/2}$ & 1.392   &   &  3.658[-6]   &    & 3.300[5]  &   & 1.500[-12]&    \\ 
\vspace{1.0mm}
$~~~~~~~~~~~~~~~~~~        \xrightarrow{E2} 3p^{5}3d^{10} \ ^2P_{3/2}$ & 0.0164  &   &  1.209[-8]   &    & 1090.48   &   &           &    \\
\vspace{1.0mm}
$~~~~~~~~~~~~~~~~~~        \xrightarrow{E1} 3p^{6}3d^9    \ ^2D_{3/2}$ & 0.108   &   &  0.149       &    & 6.671[11] &   &           &    \\

\textbf{Cd$^{21+}$} & & \\
\vspace{1.0mm}
$3p^{6}3d^9    \ ^2D_{3/2} \xrightarrow{M1} 3p^{6}3d^9    \ ^2D_{5/2}$ & 2.467   &2.391   &  9.611[-7]   &5.588[-6]    & 3213.74   &3111   & 3.111[-4] &3.21[-4]    \\
\vspace{1.0mm}       
$~~~~~~~~~~~~~~~~~~        \xrightarrow{E2} 3p^{6}3d^9    \ ^2D_{5/2}$ & 0.0047  &   &  2.545[-11]  &    & 0.0851    &   &           &    \\

\vspace{1.0mm}       
$3p^{5}3d^{10} \ ^2P_{3/2} \xrightarrow{E1} 3p^{6}3d^9    \ ^2D_{3/2}$ & 0.0208  &   &  0.026       &    & 4.473[10] &   & 2.002[-12]&    \\ 
\vspace{1.0mm} 
$~~~~~~~~~~~~~~~~~~        \xrightarrow{E1} 3p^{6}3d^9    \ ^2D_{5/2}$ & 0.191   &   &  0.162       &    & 4.545[11] &   &           &    \\

\vspace{1.0mm}       
$3p^{5}3d^{10} \ ^2P_{1/2} \xrightarrow{M1} 3p^{5}3d^{10} \ ^2P_{3/2}$ & 1.386   &   &  4.055[-6]   &    & 4.529[5]  &   & 1.390[-12]&     \\ 
\vspace{1.0mm}
$~~~~~~~~~~~~~~~~~~        \xrightarrow{E2} 3p^{5}3d^{10} \ ^2P_{3/2}$ & 0.0145  &   &  1.479[-8]   &    & 1652.60   &   &           &     \\
\vspace{1.0mm}
$~~~~~~~~~~~~~~~~~~        \xrightarrow{E1} 3p^{6}3d^9    \ ^2D_{3/2}$ & 0.102   &   &  0.147       &    & 7.196[11] &   &           &     \\

\end{tabular}
}
\end{center}
\end{ruledtabular}
\begin{tabular}{lc}
References: $^v$ \cite{Guo},  \\
\end{tabular}
\end{table*}

 In the Fock-space approach of RCC theory, we define a new working reference state $|\Phi_a \rangle = a_a | \Phi_0 \rangle$ with $a_a$ representing
annihilation operator for an the electron in the core orbital $a$ to obtain the desired reference states of our interest. Then, the exact atomic 
states are obtained by expressing \cite{Nandy3,Nandy4}
\begin{eqnarray}
|\Psi_a \rangle &=& a_a |\Psi_0 \rangle + R_a a_a |\Psi_0 \rangle \nonumber \\
                &=& \left \{ 1 + R_a \right \} e^T |\Phi_a \rangle,   
\end{eqnarray}
where $R_a$ denotes additional RCC operator that is introduced to remove the extra electron correlation effects incorporated in the determination 
of $|\Psi_0 \rangle$ due to the core electron $a$ to give rise to $|\Psi_a \rangle$. Therefore, by choosing core orbital $a$ as $3p_{3/2}$, 
$3p_{1/2}$, $3d_{3/2}$ and $3d_{5/2}$ from the configuration $[3p^{6} \ 3d^{10}]$, we can obtain the interested states of the Co-like ions using 
the above method. The amplitudes of the RCC operators $R_a$ and energy of the resulting state are obtained using the following equations  
\begin{eqnarray}
\langle \Phi^*_a| (\widehat{H_N e^T}-\Delta E_a) R_a |\Phi_a \rangle 
=  -\langle \Phi^*_a|\widehat{H_N e^T} |\Phi_a \rangle, \label{Ra_eqn} 
\end{eqnarray}
and
\begin{eqnarray}
\langle \Phi_a| \widehat{H_N e^T} \{1+R_a\}|\Phi_a\rangle = \Delta E_a \label{Eng_dett} ,
\end{eqnarray}
respectively, where $|\Phi^*_a \rangle$ corresponds to excited Slater determinants with respect to $|\Phi_a \rangle$ and $\Delta E_a = E_a - E_0$
(ionization potential (IP)) for the energy value $E_a$ of the state $|\Psi_a \rangle$. It is evident from the above two equations that they are coupled to each other and therefore, need to be solved simultaneously by adopting self-consistent procedure. Also, by taking the differences between the $\Delta E_a$ values of different states, their excitation energies (EEs) can be evaluated. Further, it is important to note that due to the choice of the DHF wave function as the starting point, the initial solution (at the first iteration) of the above two equations will correspond to the results for the second-order RMBPT (RMBPT(2)) method.

In our calculations, we have considered only the dominant singles and doubles excitations in the RCC theory (RCCSD method) by defining
$T=T_1 +T_2$ and $R_a = R_{1a}+ R_{2a}$, where and subscripts and 1 and 2 denote for the singles and doubles respectively. To make use of the 
normal ordering and Wick's theorem to reduce the amount of computation, these RCC operators are defined using the second quantization operators as
\begin{eqnarray}
   T_1 &=& \sum_{a,p}a^{\dagger}_p a_a t^p_a, 
 \ \ \ T_2 =\frac{1}{4}\sum_{ab,pq}a^{\dagger}_pa^{\dagger}_qa_ba_a t^{pq}_{ab}, \nonumber \\
   R_{1a} &=& \sum_{b\ne a}a^{\dagger}_b a_a r^b_a, 
\ \ \text{and} \ \ R_{2a} =\frac{1}{2}\sum_{bd,p}a^{\dagger}_ba^{\dagger}_pa_d a_a r^{bp}_{ad}, \ \ \ \ \
\label{Ra_quant2}
\end{eqnarray}
where the indices $a,b$ and $p,q$ represent for the core and virtual orbitals, respectively, $t$s are the amplitudes for the $T$ operators and
$r$s are the amplitudes of the $R_a$ operators.

Once atomic wave functions of the considered states of the Co-like ions are evaluated, transition matrix element due to an operator $O$ between 
the $|\Psi_f \rangle$ and $|\Psi_i \rangle$ states are determined by
\begin{eqnarray}
\frac{\langle \Psi_f | O | \Psi_i \rangle}{\sqrt{\langle \Psi_f|\Psi_f\rangle \langle \Psi_i|\Psi_i\rangle}} &=& \frac{\langle \Phi_f | \{ 1+ R_f^{\dagger}\} \overline{O}
\{ 1+ R_i\} |\Phi_i\rangle}{ \sqrt{ {\cal{N}}_f {\cal{N}}_i  }}, \nonumber \\
\label{EqnProp}
\end{eqnarray}
where $\overline{O}=(e^{T^{\dagger}} O e^T)_l$ and ${\cal N}_k = \{ (1+R_k^{\dagger}) \overline{\cal{N}} (1+R_k) \}$, where the index $k = i$ and 
$f$, with $\overline{\cal N} = (e^{T^{\dagger}} e^T)_l$, for the subscript $l$ meaning only the linked terms are contributing. It can be noted 
that the expectation value of the operator $O$ can be estimated by considering both the initial and final wave functions as same in the above 
expression. In our earlier works (e.g. see Refs. \cite{Nandy3,Nandy4}), we have discussed in detail the procedures to evaluate these terms. For 
better understanding of various contributions to the matrix elements, we explicitly quote the contributions from the normalizations of the wave 
functions using the following expression
\begin{eqnarray}
norm &=& \left [ \frac{\langle \Psi_f | O | \Psi_i \rangle} {\sqrt{\langle \Psi_f|\Psi_f\rangle
\langle \Psi_i|\Psi_i\rangle}} - \langle \Psi_f | O | \Psi_i \rangle \right ] \nonumber \\
  &=& \left [ \frac{1}{ \sqrt{ {\cal{N}}_f {\cal{N}}_i  }} - 1 \right ] \langle \Psi_f | O | \Psi_i \rangle .
\label{eqn26}
\end{eqnarray}

\section{\label{prop} Atomic properties of our interest}

\subsection{Lifetime of atomic states}
The spontaneous transition probabilities of a transition $| \Psi_i \rangle \rightarrow | \Psi_f \rangle$ due to the E1, electric-quadrupole (E2) 
and M1 channels are given by \cite{Johnson} 
\begin{eqnarray}
&& A^{E1}_{i \rightarrow f} = \frac{2.0261\times 10^{-6}}{\lambda_{ik}^3 g_i} S_{if}^{E1}, \label{eqn5} \\
&& A^{E2}_{i \rightarrow f} = \frac{1.1195\times 10^{-22}}{\lambda_{ik}^5 g_i} S_{if}^{E2} \label{eqn6} \\
\text{and}  && \nonumber \\
&& A^{M1}_{i \rightarrow f} =  \frac{2.6971\times 10^{-11}}{\lambda_{ik}^3 g_i} S_{if}^{M1}, \label{eqn7} \ \ \ \ \ \ \
\end{eqnarray}
respectively, where the quantity $S^O_{if} = \mid {\langle \Psi_f \vert \vert O \vert \vert \Psi_i \rangle} \mid^2$ is the square of the reduced matrix element between the two states with $O$ representing the corresponding E1, E2 or M1 transition operator. This is commonly known as the line strength of the electromagnetic transition and here, we calculate them in a.u.. The transition wavelength $\lambda_{if}$ used in the above formulas are taken in $cm$ and $g_i=2J_i+1$ is the degeneracy factor of the initial state $| \Psi_i \rangle$ with the angular momentum $J_i$. Thus, the transition probabilities determined using these formulas are finally given in $s^{-1}$.

\begin{table*}[t]
\caption{\label{Hyf1} The calculated ratios $A_{hf}/g_I$ (in MHz) and $B_{hf}/Q_I$ (in MHz/$b$) of the atomic $3p^{6}3d^9 \ ^2D_{5/2}$,
$3p^{6}3d^9  \ ^2D_{3/2}$, $3p^{5}3d^{10}  \ ^2P_{3/2}$ and $3p^{5}3d^{10}  \ ^2P_{1/2}$ states of the Y$^{12+}$, Zr$^{13+}$, Nb$^{14+}$, Mo$^{15+}$, Tc$^{16+}$, Ru$^{17+}$, Rh$^{18+}$, Pd$^{19+}$, Ag$^{20+}$ and Cd$^{21+}$ ions using the DHF and RCCSD methods. The $B_{hf}/Q_I$ of the Y$^{12+}$, Ag$^{20+}$ and Cd$^{21+}$ ions are not given as $B_{hf}$ of these states will be zero owing to their nuclear spin $I=1/2$.}
\begin{ruledtabular}
 %\begin{center}
 \scalebox{0.94}{
 \begin{tabular}{lcccccccccccccccc}
 %\hline \\
 \vspace{1.3mm} \\
 Ion  &   &   &   &   & $\frac{A_{hf}}{g_I}$ &   &   &   &    &    &  &  &$\frac{B_{hf}}{Q_I}$   \\ 
 \cline{2-9}  \cline{11-16} \\
 \vspace{2.0mm}
  &  \multicolumn{2}{c}{$3p^{6}3d^9   \ ^2D_{5/2}$}   &\multicolumn{2}{c}{$3p^{6}3d^9   \ ^2D_{3/2}$}  &\multicolumn{2}{c}{$3p^{5}3d^{10}  \ ^2P_{3/2}$} &\multicolumn{2}{c}{$3p^{5}3d^{10}   \ ^2P_{1/2}$}&  &\multicolumn{2}{c}{$3p^{6}3d^9   \ ^2D_{5/2}$}  &\multicolumn{2}{c}{$3p^{6}3d^9   \ ^2D_{3/2}$}  &\multicolumn{2}{c}{$3p^{5}3d^{10}  \ ^2P_{3/2}$}\\
  &DHF  & RCCSD   &DHF   & RCCSD   &DHF   & RCCSD  &DHF  & RCCSD  &  &DHF  &RCCSD  &DHF  &RCCSD  & DHF & RCCSD \\
 \cline{2-3} \cline{4-5} \cline{6-7} \cline{8-9} \cline{11-12} \cline{13-14} \cline{15-16} \\
 \vspace{2.0mm}  
 
 \text{Y$^{12+}$} &2651 &2753 &6331  &6904 &14925 &16355 &86147  &93936  & & \\
 \vspace{1.8mm}
 \text{Zr$^{13+}$}&2991 &3102 &7151  &7765 &16587 &18077 &96486  &104668 & &6160 &6242 &4492 &4555  &31357&33487 \\
 \vspace{1.8mm}
 \text{Nb$^{14+}$}&3361 &3477 &8036  &8693 &18375 &19932 &107642 &116271 & &6919 &7000 &5056 &5118  &34786&37007 \\
 \vspace{1.8mm}
 \text{Mo$^{15+}$}&3752 &3881 &8992  &9691 &20291 &21921 &119794 &128922 & &7735 &7813 &5665 &5726  &38461&40782 \\
 \vspace{1.8mm}
 \text{Tc$^{16+}$}&4176 &4313 &10021 &10762&22336 &24044 &132940 &142607 & &8611 &8686 &6321 &6380  &42400&44825 \\
 \vspace{1.8mm}
 \text{Ru$^{17+}$}&4631 &4775 &11121 &11911&24517 &26308 &147227 &157478 & &9548 &9618 &7026 &7082  &46604&49141 \\
 \vspace{1.8mm}
 \text{Rh$^{18+}$}&5113 &5267 &12295 &13133&26842 &28721 &162616 &173490 & & & & &  && \\
 \vspace{1.8mm}
 \text{Pd$^{19+}$}&5627 &5791 &13551 &14441&29314 &31285 &179231 &190771 & &11614&11672&8591 &8638  &55887&58660 \\
 \vspace{1.8mm}
 \text{Ag$^{20+}$}&6174 &6346 &14891 &15831&31935 &34004 &197381 &209651 & & \\
 \vspace{1.8mm}
 \text{Cd$^{21+}$}&6756 &6941 &16312 &17304&34724 &36892 &216274 &229276 & & \\
 %
 %\hline                 
 \end{tabular}
 }
 %\end{center}
\end{ruledtabular}
\end{table*}

Another, useful quantity which could be of particular interest in the astrophysical study is the emission (absorption) oscillator strengths 
$F_{if}$ ($F_{fi}$). This quantity can be deduced from the above transition probabilities through the following expressions \cite{Sobelman}
\begin{eqnarray}
F_{if}^O = 1.4992\times 10^{-24} A_{if}^O \frac{g_i}{g_f}\lambda_{if}^2 ,
\label{eqn3}
\end{eqnarray}
which follows that $g_f F_{fi}^O = - g_i F_{if}^O$.

The lifetime of a given atomic state is the inverse of the total transition probabilities involving all possible spontaneous emission channels; 
i.e. the lifetime (in $s$ corresponding to the units used above) of the state $|\Psi_f \rangle$ is given by
\begin{eqnarray}
\tau_i &=& \frac {1} {\sum_{O,f} A^{O}_{i\rightarrow f}},
\label{life}
\end{eqnarray}
where sum over $O$ represents all possible decay channels due to transition operators $O$ and the summation index $f$ corresponds to all the final atomic states. 

\subsection{Hyperfine interaction coefficients}

The Hamiltonian describing the non-central form of hyperfine interaction between the electrons and nucleus in an atomic system is expressed in
terms of spherical tensor operator products as \cite{Schwartz, Lindgren}
\begin{eqnarray}
H_{hf} = \sum_k {\bf M}_n^{(k)} \cdot {\bf O}_{hf}^{(k)},
\end{eqnarray}
where ${\bf M}_n^{(k)}$ and ${\bf O}_{hf}^{(k)}$ are the spherical tensor operators with rank $k$ ($>0$) in the nuclear and electronic coordinates
respectively. Since these interaction strengths become much weaker with higher values of $k$, we consider only up to $k=2$ for the present interest.
Also, we account only the first-order effects due to these interactions giving rise to the energy shift to an energy level
\begin{eqnarray}
W_{F,J} &=& \langle H_{hf} \rangle = A_{hf} {\bf I.J} \nonumber \\
&& + B_{hf} \frac {3({\bf I.J})^2 + \frac {3}{2}({\bf I.J})- I(I+1)J(J+1)}{2I(2I-1)J(2J-1)} , \ \ \ \ \ \ 
\end{eqnarray}
where $I$ and $J$ are the nuclear and atomic angular momenta, respectively, and $A_{hf}$ and $B_{hf}$ are known as the M1 and E2 hyperfine 
structure constants. With the knowledge of $A_{hf}$ and $B_{hf}$, it is possible to estimate $W_{F,J}$ for any hyperfine level $F=I+J$.
Thus, we evaluate these constants using the expressions 
\begin{eqnarray}
A_{hf}&=& \mu_N g_I \frac {\langle J||O_{hf}^{(1)}||J\rangle}{\sqrt{J(J+1)(2J+1)}} \label{eqnA} 
\end{eqnarray}
and
\begin{eqnarray}
B_{hf}&=& 2 Q_I [ \frac {2J(2J-1)}{(2J+1)(2J+2)(2J+3)}]^{1/2} \nonumber \\ && \times \langle J||O_{hf}^{(2)}||J\rangle \label{eqnB},
\end{eqnarray}
where $\mu_N$ is the nuclear Bohr magneton, $g_I= \frac {\mu_I}{I}$, $\mu_I$ and $Q_I$ are the nuclear M1 and E2 moments respectively. Since 
the $A_{hf}/g_I$ and $B_{hf}/{Q_I}$ values are independent of isotopes and depend only on the atomic wave functions, determination of these 
quantities are our particular interest.

\begin{figure*}[t]
\includegraphics[width=17.0cm, height=6.5cm, clip=true]{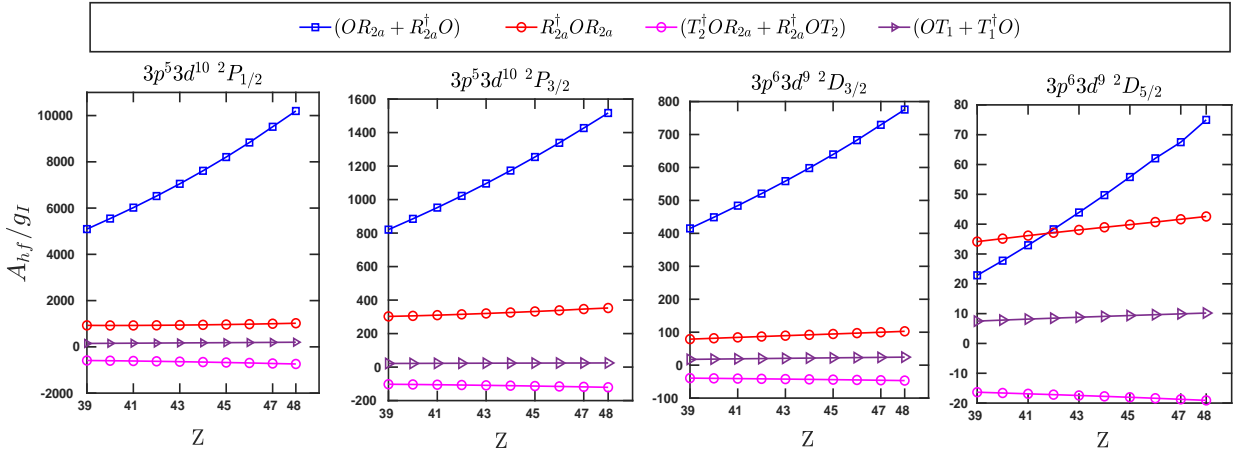}
\caption{(color online) Plots showing the contributions from different dominant RCC terms such as ($OR_{2a} + R_{2a}^{\dagger}O$), 
($OT_1 + T_1^{\dagger}O$), $R_{2a}^{\dagger}OR_{2a}$ and ($R_{2a}^{\dagger}OT_{2} + T_2^{\dagger}OR_{2a}$) in the calculations of the ratios 
$A_{hf}/g_I$ (in MHz) for the calculated states against the atomic number $Z$ of the considered ions.} 
\label{fig_Ahyf}
\end{figure*} 

\section{\label{Results} Results and Discussion}

As mentioned earlier, we calculate first the ground state configurations of the ions having Ni isoelectronic sequence and then, atomic state of 
the Co-like ions are determined by removing an electron from the occupied orbitals of the Ni-like ions. In this process, we obtain the first IPs 
of the respective Ni-like ions. However, the differential values between the IPs of different orbitals correspond to the EEs of the Co-like ions. 
The calculated IPs for the electrons in the $d_{5/2}$, $d_{3/2}$, $p_{3/2}$ and $p_{1/2}$ orbitals giving rise to the $3p^6 3d^{9}  \ ^2D_{5/2}$, 
$3p^6 3d^{9} \ ^2D_{3/2}$, $3p^5 3d^{10} \ ^2P_{3/2}$ and $3p^5 3d^{10} \ ^2P_{1/2}$ atomic states of the investigated Co-like ions are given in 
the Table \ref{IPs} from the DHF, RMBPT(2) and RCCSD methods. Contributions from the leading order relativistic corrections such as Breit 
interaction ($\Delta E_B$), VP effect ($\Delta E_{VP}$) and SE effect ($\Delta E_{SE}$) are also estimated and quoted in the above table 
explicitly. From these tabulated values for IPs, we find after the Coulomb interactions the Breit interactions also contribute significantly to 
the energy. There are large cancellations among the VP and SE effects of the QED interactions. These IP values also show that the DHF method 
overestimates the energies, while there is a gradual decrease in the values from the RMBPT(2) to RCCSD methods using the DC Hamiltonian. Further 
analysis demonstrates that contributions from the correlation and the relativistic effects are increasing from the ground state to the excited 
states. The trends of the correlation effects are found to be similar in all the considered Co-like ions using our RCC theory.    

We present the final values of the IPs of all the four low-lying states for the investigated ions in Table  \ref{IPs} by adding contributions from
the DC Hamiltonian and corrections from the Breit, VP, and SE interactions. We have also estimated uncertainties to the total values by analyzing 
contributions due to the truncation of basis functions and neglected higher-level excitations in the RCC theory. The basis function extrapolations
are obtained using a lower-order many-body method while we have estimated uncertainties due to the higher level excitations by analyzing 
contributions from the dominant triple excitations by adopting the perturbative approach. Our final values are also compared with the IPs of the 
only available data for the $3p^6 3d^{9}  \ ^2D_{5/2}$ states for all the ions from the National Institute of Science and Technology (NIST) 
database \cite{NIST}. These values were obtained using the non-relativistic Hartree-Fock orbitals, so we see large differences among these values. Nonetheless, IPs for the orbitals giving rise to the other states of Co-like ions are not available for comparison. 

It can be obvious from the above discussions on IPs that EEs, which are obtained from the differences of IPs, are more relevant quantities here as
they are directly related to the investigated Co-like ions. In Table \ref{EEs}, we compare our calculated EEs with a few available experimental 
results. Only a few direct measurements of excitation energies are reported, while the other experimental values are extrapolated by fitting the 
calculated wavelengths with some of the observed wavelengths. So far, the direct measurements were carried out only for the ions Zr$^{13+}$, 
Mo$^{15+}$, and Nb$^{14+}$. Edl\'{e}n \cite{Edlen} had measured the forbidden lines of the Zr$^{13+}$ and Mo$^{15+}$ ions in a hot tokamak plasma 
experiment, while Prior \cite{Prior} had directly obtained the EEs of Nb$^{14+}$ by performing measurement using the electron cyclotron resonance 
ion source. The indirectly inferred values are quoted in the above table as `Fitted', which had used calculations using the MCDHF method to 
extrapolate EEs of all the considered ions \cite{Ekberg}. Comparison between our calculated values with the measurements shows good agreement
between them suggesting our calculations for the transition matrix elements using the RCC theory can be accurate enough to estimate the transition properties of the excited states. This also suggests that the inclusion of triple excitations in our RCC calculations can improve our results further. 

\begin{figure*}[t]
\includegraphics[width=14.0cm, height=6.5cm, clip=true]{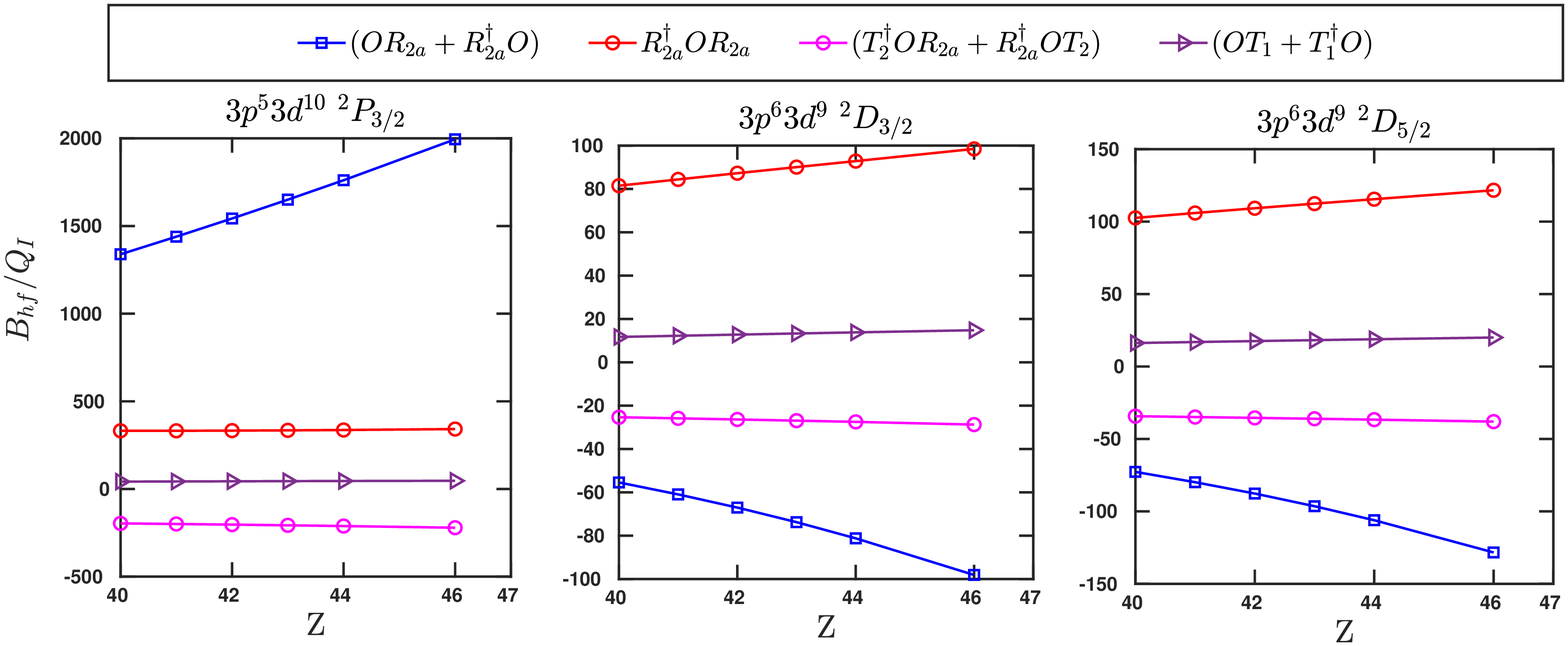}
\caption{(color online) Plots showing the contributions from different dominant RCC terms such as ($OR_{2a} + R_{2a}^{\dagger}O$), 
($OT_1 + T_1^{\dagger}O$), $R_{2a}^{\dagger}OR_{2a}$ and ($R_{2a}^{\dagger}OT_{2} + T_2^{\dagger}OR_{2a}$) in the evaluation of the ratios 
$B_{hf}/Q_I$ (in MHz) against the atomic number $Z$ of the Zr$^{13+}$, Nb$^{14+}$, Mo$^{15+}$, Tc$^{16+}$, Ru$^{17+}$, and Pd$^{19+}$ ions.
Since, $I = 1/2$, for the ions \text{$^{89}_{39}$Y$^{12+}$}, \text{$^{103}_{45}$Rh$^{18+}$}, \text{$^{109}_{47}$Ag$^{20+}$}, and 
\text{$^{111}_{48}$Cd$^{21+}$}, so the $Q_{I}$ value for them do not exist. This is why we have excluded the calculation of the
$B_{hf}/Q_I$ ratios for these ions.} 
\label{fig_Bhyf}
\end{figure*} 

After analyzing the accuracies of the calculated EEs using our RCCSD method, we now proceed to calculate other transition properties such as the
line strengths, transition probabilities, oscillator strengths, and lifetimes of the excited states of the considered Co-like ions. We also present
the hyperfine structure constants of all the calculated states. The transition properties such as the line strengths, oscillator strengths, 
transition probabilities, and lifetimes of the excited states are presented in Table \ref{Prob}. First, the line strengths are determined using 
the calculated reduced matrix elements of the E1, E2, and M1 operators. Substituting these values, we obtained the other transition properties. In
order to reduce the uncertainties, we have used the wavelengths from the NIST database in estimating these values. Earlier, lifetimes of the fine 
structure level of the ground state of the aforementioned ions were estimated by applying the MCDHF method, and we found reasonable agreement 
among our values with the previously estimated values. In the earlier estimations, contributions from the E2 channel were neglected and our 
analysis shows that they are indeed small. The lifetimes of the $3p^{5}3d^{10} \ ^2P_{3/2}$ and $3p^{5}3d^{10} \ ^2P_{1/2}$ states are not 
available to date, so we are unable to make a comparative analysis of these values. In the determination of lifetimes of the $3p^{5}3d^{10} \ 
^2P_{1/2}$ states, we have also accounted for the transition probabilities due to the forbidden channels but their contributions are found to be 
negligibly small compared to the E1 probability contributions. The E1 transition probabilities of the $3p^{5}3d^{10} \ ^2P_{3/2} \rightarrow 
3p^{6}3d^9 \ ^2D_{5/2}$ transitions are found to be dominant over the $3p^{5}3d^{10} \ ^2P_{1/2} \rightarrow 3p^{6}3d^9 \ ^2D_{5/2}$ transitions. Though there are two E1 transitions are allowed from the $3p^{5}3d^{10} \ ^2P_{3/2}$ state than the $3p^{5}3d^{10} \ ^2P_{1/2}$, the lifetimes of the $3p^{5}3d^{10} \ ^2P_{1/2}$ states in the Co-like ions are found to be smaller than the $3p^{5}3d^{10} \ ^2P_{3/2}$ states. We also find that the E1 transition probabilities are larger when the angular momentum difference is $|\Delta J=1|$ than $|\Delta J=0|$. Further, due to the monotonic 
increase in the energy gap between the $3p^{5}3d^{10} \ ^2P_{3/2}$ and the $3p^{6}3d^{9} \ ^2D_{5/2}$ ground state with the size of the ion, the 
transition probabilities gradually increase from Y$^{12+}$ to Tc$^{16+}$. This results in smaller values of the lifetimes of the atomic states 
with increasing ionic charge of the Co-like systems.     
  
\begin{table*}[t]
 \caption{\label{Hyf2} The estimated values of $A_{hf}$ and $B_{hf}$ for the calculated states of the Co-like ions using the RCCSD method.
The nuclear parameters for the stable isotopes used to estimate these values are taken from Ref. \cite{Stone} and they are listed here. As can be 
seen the reported $Q_I$ values of the $^{91}_{40}$Zr$^{13+}$ and $^{97}_{42}$Mo$^{15+}$ isotopes differ significantly from various works, so we present 
the $B_{hf}$ values for these ions by considering all the reported values of $Q_I$. We anticipate that the $Q_I$ values of these isotopes can be 
inferred more reliably by combining our calculations with possible measurements of the $B_{hf}$ values in these ions.}
 \begin{ruledtabular}
 %\begin{center}
 %\scalebox{0.90}{
 \begin{tabular}{lccccccccccccc}
 %\hline \\
 \vspace{1.3mm} \\
 Ion   &$I$   &$\mu_I$  & $g_I$ &  & $A_{hf}$ (in MHz)   &   &   &    &$Q_I$ (in $b$) &   &$B_{hf}$ (in MHz)    &   & \\ 
 \cline{5-8}  \cline{11-13} \\
 \vspace{2.0mm}
  & &  & & \multicolumn{1}{c}{$^2D_{5/2}$}  &\multicolumn{1}{c}{$^2D_{3/2}$} & \multicolumn{1}{c}{$\ ^2P_{3/2}$}  &\multicolumn{1}{c}{$^2P_{1/2}$} &  &   &  \multicolumn{1}{c}{$^2D_{5/2}$} & \multicolumn{1}{c}{$^2D_{3/2}$} & \multicolumn{1}{c}{$^2P_{3/2}$} \\
 %\hline \\
 \cline{5-8}  \cline{11-13} \\
 \vspace{3.0mm}  
 \text{$^{89}_{39}$Y$^{12+}$} & $\frac{1}{2}$ & $-0.1374154(3)$&$-0.2748308$ &$-756$  &$-1897$   &$-4494$  &$-25816$  &  &  & & & \\

 \vspace{3.0mm}
 \text{$^{91}_{40}$Zr$^{13+}$}& $\frac{5}{2}$ & $-1.30362(2)$  &$-0.521448$ &$-1617$  &$-4049$   &$-9426$  &$-54578$  &  & $-0.176(3)$ 
 &$-1098$   &$-801$   &$-5893$  \\                        
 \vspace{2.0mm}
                              &                   &                & &        &         &         &          &  & $-0.257(13)$&$-1604$    &$-1170$   &$-8606$     \\ 
                                                      
 \vspace{2.0mm}
                              &                   &                & &        &         &         &          &  & $-0.206(10)$&$-1285$    &$-938$    &$-6898$    \\ 
                                                      
 \vspace{3.0mm}
 \text{$^{93}_{41}$Nb$^{14+}$}& $\frac{9}{2}$ & $+6.1705(3)$   &$1.37122$ &$4767$  &$11920$   &$27331$  &$159433$  &  & $-0.37(2)$  
 &$-2590$ &$-1893$ &$-13692$ \\
 \vspace{3.0mm}
 \text{$^{97}_{42}$Mo$^{15+}$}& $\frac{5}{2}$ & $-0.9335(1)$   &$-0.37340$ &$-1449$  &$-3618$   &$-8185$  &$-48139$  &  & $0.255(13)$
 &$1992$   &$1460$   &$10399$  \\
 \vspace{2.0mm}
                              &                   &                & &        &         &         &          &  & $0.17(4)$  &$1328$ &$973$ &$6932$    \\
 
 \vspace{2.0mm}
                              &                   &                & &        &         &         &          &  & $0.27(10)$  &$2109$ &$1546$ &$11011$    \\ 
 \vspace{3.0mm}
 \text{$^{99}_{43}$Tc$^{16+}$}& $\frac{9}{2}$ & $+5.6847(4)$   &$1.263266$ &$5448$  &$13595$  &$30374$  &$180150$  &  & $-0.129(6)$ 
 &$-1120$ &$-823$  &$-5782$  \\
 \vspace{3.0mm}
 \text{$^{101}_{44}$Ru$^{17+}$}&$\frac{5}{2}$ & $-0.719(6)$    &$-0.28760$ &$-1373$  &$-3425$  &$-7566$  &$-45290$  &  & $0.46(2)$  
 &$4424$   &$3257$   &$22605$  \\
 \vspace{3.0mm}
 \text{$^{103}_{45}$Rh$^{18+}$}& $\frac{1}{2}$& $-0.8840(2)$   &$-1.7680$  &$-9312$  &$-23219$  &$-50779$  &$-306730$  & & &  & &  \\
 \vspace{3.0mm}
 \text{$^{105}_{46}$Pd$^{19+}$}& $\frac{5}{2}$& $-0.642(3)$    &$-0.25680$ &$-1487$  &$-3708$  &$-8034$  &$-48990$  &  & $0.660(11)$ 
 & $7703$ & $5701$  & $38715$ \\
 \vspace{2.0mm}
                              &                   &                & &        &         &         &          &  & $0.65(3)$   &$7587$  &$5615$ 
                              &$38129$    \\ 
 \vspace{3.0mm}
 \text{$^{109}_{47}$Ag$^{20+}$}& $\frac{1}{2}$& $-0.1306906(2)$&$-0.2613812$ &$-1659$  &$-4138$  &$-8888$  &$-54799$  & & &  &   &  \\
 \vspace{3.0mm}
 \text{$^{111}_{48}$Cd$^{21+}$}& $\frac{1}{2}$& $-0.5948861(8)$&$-1.1897722$ &$-8258$  &$-20588$  &$-43893$  &$-272786$  &  &  &  &  &  \\
 %
 %\hline                 
 \end{tabular}
 %}
 %\end{center}
 \end{ruledtabular}
 \end{table*}
 
Now we turn on to present the results for the hyperfine structure constants of the considered Co-like ions. The accuracies of the transition 
matrix elements discussed earlier depend on the accurate determinations of the wave functions in the asymptotic region while accuracies in the 
evaluation of the hyperfine structure constants depend on the accurate calculations of the wave functions in the nuclear region. The determination of the hyperfine structure constants not only depends on the accurate calculations of the atomic matrix elements but also requires knowledge of accurate values of the nuclear moments. Since we are interested to estimate the $A_{hf}$ and $B_{hf}$ values, we need prior knowledge of $g_I=\frac{\mu_I}{I}$ and $Q_I$ of the isotopes of the interest. This implies the $A_{hf}$ and $B_{hf}$ values are isotope dependent. However, the calculations of the $A_{hf}/g_I$ and $B_{hf}/Q_I$ values hardly change with the nuclear structure of the isotopes of an element. Thus, we discuss first these results and then present the estimated $A_{hf}$ and $B_{hf}$ value only for the stable isotopes of the elements of the investigated Co-like ions by combining with their respective $g_I$ and $Q_I$ values. Our calculated values of $A_{hf}/g_I$ and $B_{hf}/Q_I$ are reported in Table \ref{Hyf1} for all the considered atomic states of the Co-like Y$^{12+}$ - Cd$^{21+}$ ions. We have not given the $B_{hf}/Q_I$ values of the Y$^{12+}$, Rh$^{18+}$, Ag$^{20+}$ and Cd$^{21+}$ ions as their $B_{hf}$ values do not exist owing to the fact that they all have $I=1/2$. It can be observed from this table that the DHF values for $A_{hf}/g_I$ are smaller than the RCCSD results for all the states, which are opposite to the trends seen in the calculations of IPs. The values and the electron correlation effects increase from the ground to the higher excited states. The reason for the large magnitude is due to the fact that the $3d$ orbitals have less overlap with the nucleus than the $3p$ orbitals, which are the valence orbitals of the first and the last two states respectively. The possible reason for which the correlation effects are seen to be enhanced in the calculations of the hyperfine structure constants for the ground state to the higher level excited states are probably due to the large correlations among the $s$ and $p$ orbitals than the $s$ and $d$ orbitals. Again, the values of the above quantities are found to be increasing with the size of the ion. The reason for this could be due to highly contracted orbitals in the more highly charged ions that can overlap with the nucleus strongly.  

We also intend to fathom the roles of different electron correlation effects in the atomic states of Co-like ions. Evaluation of transition matrix
elements depends on the wave functions of two different atomic states, while the determination of hyperfine structure constants of a state depends 
only on the wave function of the respective state. Thus, we analyze the contributions to the $A_{hf}/g_I$ and $B_{hf}/Q_I$ values arising through 
various RCCSD terms. Instead of quoting them in tables, we show their contributions to the $A_{hf}/g_I$ and $B_{hf}/Q_I$ values in the graphical 
representations in Figs. \ref{fig_Ahyf} and \ref{fig_Bhyf}, respectively, against the atomic number. Among all property evaluating RCC terms, we
find that the $OR_{2a}$, $OT_1$, $R_{2a}^{\dagger}OR_{2a}$ and $R_{2a}^{\dagger}OT_{2}$ terms along with their hermitian conjugate (h.c.) 
contribute predominantly to the above quantities. The term representing $OR_{2a}$ accounts for the core-polarization effects to all-orders, while 
the $OT_1$ term represents for the extra core-valence correlation effects that were accounted in the calculations of the ground states of the 
corresponding Ni-like ions from which atomic states of the Co-like ions were derived. The other two non-linear terms, $R_{2a}^{\dagger}OR_{2a}$ 
and $R_{2a}^{\dagger}OT_{2}$, are responsible for including higher-order core-polarization effects in our calculations. It can be seen from  
Fig. \ref{fig_Ahyf} that the most dominating term is the core-polarization term $OR_{2a}$ for all the atomic states that further show an 
increasing trend with atomic number. As expected, the effect of the core-polarization for the outermost $d-$orbitals are comparatively quite 
smaller than the inner valence $p-$orbitals, so the contribution to the $A_{hf}/g_I$ values are quite large for the $3p^{5}3d^{10} \ ^2P_{1/2, 3/2}$
excited states. The next dominating contribution comes from the non-linear term $R_{2a}^{\dagger}OR_{2a}$ although the magnitude is smaller 
compared to the core-polarization effect except for the ground states with $Z = 39$, $40$ and $41$. The other non-linear term, $R_{2a}^{\dagger}
OT_{2}$, also contributes significantly however, the values show an opposite behavior (i.e. negative value) compared to the other three terms. 
Finally, the core-valence correlation effects through $OT_1$ seem to give non-negligible contribution to $A_{hf}/g_I$. 

We now would like to discuss the behavior of the above dominating terms for the calculations of $B_{hf}/Q_I$ and the contributions from the above RCC terms to this quantity are plotted in Fig. \ref{fig_Bhyf}. The behavior for the core-polarization effect in determining the $B_{hf}/Q_I$ values are found to be quite similar to that of $A_{hf}/g_I$ for the excited state $3p^{5}3d^{10} \ ^2P_{3/2}$ although they differ in the magnitudes percentage-wise.  In contrast, for the ground state doublets, the core-polarization trend shows an opposite behavior as compared to the $A_{hf}/g_I$ values for the excited states. In fact, it shows an increasing trend in the negative direction with respect to the atomic number. The next leading order contributions to $B_{hf}/Q_I$ are given by the $R_{2a}^{\dagger}OR_{2a}$ term which further show that for the state $3p^{5}3d^{10} \ ^2P_{3/2}$ their magnitudes are nearly equal for all the investigated ions. On contrary, for the ground state doublets, the corresponding values are slowly 
increasing as a function of $Z$. There are also finite contributions coming from the non-linear term $R_{2a}^{\dagger}OT_{2}$ that show an almost 
constant trend in the respective states with the increase in atomic number. The core-valence term $OT_1$ also gives non-negligible contributions 
to the $B_{hf}/Q_I$ values for all the states. 

As mentioned earlier, the quantities of experimental interest are the $A_{hf}$ and $B_{hf}$ values. To obtain these values from our calculations of
$A_{hf}/g_I$ and $B_{hf}/Q_I$, we used the nuclear moments that are listed in the nuclear data table \cite{Stone} for the most stable isotopes.
We have given the final $A_{hf}$ and $B_{hf}$ values for all the four calculated states by combining our RCCSD values of atomic calculations and 
nuclear moments in Table \ref{Hyf2}. Due to the fact that $I=1/2$, the $B_{hf}$ values do not exist for Y$^{12+}$, Rh$^{18+}$, Ag$^{20+}$ and 
Cd$^{21+}$. The nuclear moments for the stable isotopes for which we have determined the hyperfine structure constants are also listed in the 
above table. It can be seen that the $\mu_I$ values are known very precisely for these isotopes, but many different $Q_I$ values are reported for a few isotopes; especially for $^{91}_{40}$Zr$^{13+}$ and  $^{97}_{42}$Mo$^{15+}$. So we suggest that if the $B_{hf}$ of either of the $3p^{6}3d^9 
\ ^2D_{5/2}$, $3p^{6}3d^9 \ ^2D_{3/2}$ or $3p^{5}3d^{10} \ ^2P_{3/2}$ state is measured precisely for the above ion, then by combining that 
measured value with our $B_{hf}/Q_I$ calculation it is possible to infer the $Q_I$ value of the respective ion more reliably.      
 
\section{Conclusion}

We have employed the Fock-space relativistic coupled-cluster method to calculate the atomic wave functions of the first four low-lying 
$3p^{6}3d^9 \ ^2D_{5/2}$, $3p^{6}3d^9 \ ^2D_{3/2}$, $3p^{5}3d^{10} \ ^2P_{3/2}$ and $3p^{5}3d^{10} \ ^2P_{1/2}$ states of the Co-like ions such as 
Y$^{12+}$, Zr$^{13+}$, Nb$^{14+}$, Mo$^{15+}$, Tc$^{16+}$, Ru$^{17+}$, Rh$^{18+}$, Pd$^{19+}$, Ag$^{20+}$ and Cd$^{21+}$, which are one electron 
less than a closed-shell electronic configuration. The Dirac-Breit interactions along with lower-order QED effects through an effective potential 
are considered to perform these calculations. Only the dominant singles and doubles excitation configurations were taken into account in our 
method, and the uncertainties were estimated by analyzing leading order contributions from the valence triple excitations and truncated basis 
functions. The ionization potentials of the Ni-like ions of the above elements were first determined in order to obtain the considered atomic 
states of Co-like ions, and taking their differences the excitation energies of the respective Co-like ions were estimated. Further, the 
calculated wave functions were used to determine the E1, E2, and M1 transition matrix elements among the aforementioned states of the Co-like 
ions. Further, using these matrix elements we determine other transition properties such as the line strengths, oscillator strengths, and 
transition probabilities. The lifetimes of the excited states were estimated from the total transition probabilities from a given excited state 
and they are compared with the available theoretical values. In addition, we have also determined the magnetic-dipole and electric-quadrupole 
hyperfine structure constants of the above states of the stable isotopes of Co-like ions. Since the nuclear quadrupole moment of the 
$^{91}_{40}$Zr and $^{97}_{42}$Mo isotopes are not known precisely, we suggest to infer their values by combining our calculations of 
$B_{hf}/Q_I$ of one of its states with the measurement of $B_{hf}$ of the corresponding state in the future.

\section*{Acknowledgment}
DKN acknowledges use of the high-performance computing facility (FERMI cluster) at IBS-PCS and BKS acknowledges use of Vikram-100 HPC facility 
for performing calculations and implementation of the program. 

%\section{References}

%\bibliographystyle{aipnum4-1}
\bibliographystyle{apsrev4-1}
\nocite{apsrev41Control}
\bibliography{CoLike}

%\bibliography{CoLike}

\end{document}